\documentclass[12pt]{iopart}
\usepackage{iopams}  
\usepackage{graphicx}
\usepackage[usenames,dvipsnames]{xcolor}
\usepackage{hyperref}
\usepackage{enumitem}

\hypersetup{
    colorlinks=true,
    linkcolor=cyan,
    filecolor=magenta,      
    urlcolor=blue,
    citecolor=blue,
}

\begin{document}

\title[Physics-inspired spatiotemporal-graph AI ensemble for gravitational wave detection]{Physics-inspired spatiotemporal-graph 
AI ensemble for the detection of higher order wave mode signals of spinning binary black hole mergers}


\author{Minyang Tian$^{1,2,3}$, E.~A. Huerta$^{1,2,4}$, Huihuo Zheng$^{5}$, and Prayush Kumar$^{6}$}
\address{$^1$ Data Science and Learning Division, Argonne National Laboratory, Lemont, Illinois 60439, USA}
\address{$^2$ Department of Physics, University of Illinois at Urbana-Champaign, Urbana, Illinois 61801, USA}
\address{$^3$ NCSA, University of Illinois at Urbana-Champaign, Urbana, Illinois 61801, USA}
\address{$^4$ Department of Computer Science, The University of Chicago, Chicago, Illinois 60637, USA}
\address{$^5$ Leadership Computing Facility, Argonne National Laboratory, Lemont, Illinois 60439, USA}
\address{$^6$ International Centre for Theoretical Sciences,
Tata Institute of Fundamental Research, Bangalore 560089, India}

\ead{mtian8@illinois.edu}


\begin{abstract}
We present a new class of AI models for 
the detection of quasi-circular, spinning, 
non-precessing binary black hole mergers 
whose waveforms include the higher order gravitational wave modes  $(\ell, |m|)=\{(2, 2), (2, 1), (3, 3), 
(3, 2), (4, 4)\}$, and mode mixing effects 
in the \(\ell = 3, |m| = 2\) harmonics.  
These AI models combine hybrid dilated convolution 
neural networks to accurately model both short- and long-range 
temporal sequential information of gravitational waves; 
and graph neural networks to capture spatial 
correlations among gravitational wave observatories 
to consistently describe and identify the presence of a signal 
in a three detector network encompassing the Advanced 
LIGO and Virgo detectors. We first trained these 
spatiotemporal-graph AI models using synthetic noise, 
using 1.2 million modeled waveforms to densely sample this 
signal manifold, within 1.7 hours using 256 NVIDIA 
A100 GPUs in the Polaris 
supercomputer at the Argonne Leadership Computing 
Facility. This distributed training approach exhibited optimal classification performance, and strong 
scaling up to 512 NVIDIA 
A100 GPUs. With these AI 
ensembles we processed data from a three detector 
network, and found that an 
ensemble of 4 AI models achieves 
state-of-the-art 
performance for signal detection, and reports two 
misclassifications for every decade of 
searched data. We distributed AI inference over 
128 GPUs in the Polaris 
supercomputer and 128 nodes in the Theta supercomputer, 
and completed the processing of a decade of gravitational 
wave data from a three detector network within 3.5 hours. 
Finally, we fine-tuned 
these AI ensembles to 
process the entire month of February 2020, 
which is part of the O3b 
LIGO/Virgo observation run, and found 6 gravitational waves, 
concurrently identified in Advanced LIGO and 
Advanced Virgo data, and zero false positives. 
This analysis was completed in one hour using one 
NVIDIA A100 GPU.
\end{abstract}

%
%
%
%
%

\section*{Introduction}

The development of AI methodologies for gravitational 
wave astrophysics is a booming enterprise. Since 2017~\cite{mma:2017_bns,2021arXiv210909882E,2021NatRP344B,2020JCAP050M}, 
there has been a rapid development of AI methods and approaches 
to create, test and deploy production scale tools for 
gravitational wave detection. These novel AI methodologies 
are being explored in earnest to provide alternatives 
to well established, though compute intensive and poorly scaling,  signal processing tools, such as template 
matching algorithms, which utilize large sets of modeled 
waveforms to search for signals in gravitational 
wave data~\cite{2016CQGra..33u5004U,CANNON2021100680}. The first class of 
AI models for gravitational wave 
detection 
were designed to describe 2D gravitational 
wave signal manifolds, 
comprising the masses, \((m_1, m_2)\), of 
non-spinning, quasi-circular 
binary black hole mergers, and which were 
capable of detecting true gravitational wave 
signals with a false positive rate of 
one misclassification for every 100 seconds of 
searched 
data~\cite{geodf:2017a,George:2017vlv,George:2018PhLB}. 
The development of these AI models 
required training datasets with a few tens 
of thousands of modeled signals, and a couple of 
inexpensive GPUs to complete the training within  
a few hours. 
These findings sparked the interest of the 
gravitational wave community, leading to the 
organic creation of a vibrant, 
international community of researchers who 
are harnessing advances in AI and computing 
to address timely and pressing challenges in 
gravitational wave astrophysics~\cite{2018GN,Lin:2020aps,Wang:2019zaj,Fan:2018vgw,Li:2017chi,Deighan:2020gtp,Adam:2018prd,2020PhRvD.101f4009B,2022PhRvD.105d3003S,2022PhRvD.105d3002S,2022NatAs...6..529G,2023PhRvD.107b3021S,Andrews:2022kfh,2021arXiv211103295Y,Moreno:2021fvp,Nemmen_Duarte_Navarro_2019,2021IJGMM..1850154O,2018EL....12450002A}. 
These seminal ideas have been applied 
for signal 
detection~\cite{Krastev:2019koe,2020PhRvD.102f3015S,Miller:2019jtp} and 
forecasting~\cite{Wei:2020sfz,2021PhRvD.104f2004Y,Qiu:2022wub} 
of binary neutron stars, and for 
the detection and forecasting of neutron 
star-black hole systems~\cite{Wei:2020sfz,Wei_quantized}. 
Comprehensive reviews of 
this active area of research may be found in 
Refs.~\cite{huerta_book,Nat_Rev_2019_Huerta,
2022hcsr.book..193B}, while Ref.~\cite{cuoco_review} provides a wider view on other problems tackled in the literature 
with machine-learning methods such as detector noise characterisation and glitch classification, core-collapse supernovae detection, parameter estimation, etc.

Promoting these disruptive ideas into production scale 
frameworks for gravitational wave discovery requires innovation 
at the interface of  
AI and supercomputing~\cite{2020arXiv200308394H}. 
This is because AI models 
that describe compact binary systems that may be 
detectable by ground-based interferometric detectors 
span a high dimensional signal manifold. Assuming 
astrophysical compact binary sources that 
spiral into each other following a series 
of quasi-circular orbits, and whose individual
components are spinning and non-precessing, 
may be described in terms of four parameters, 
\((m_1, m_2, s_1^z,s_2^z)\). This 4D signal manifold 
needs to be smartly and densely sampled to train AI models so 
as to capture the physics of these gravitational 
wave sources. This then translates into training 
datasets that have tens of millions 
of modeled waveforms, i.e., Terabyte size datasets. 
Therefore, in order to reduce time-to-solution, 
it is critical to use distributed training algorithms 
that optimally utilize between hundreds to thousands 
of GPUs in supercomputing platforms. 
Furthermore, these methodologies enable the design 
of AI surrogates that 
incorporate physics and math principles 
in their architecture, training and optimization. 
In parallel to these development, 
it is essential to create new methods to improve the 
sensitivity and performance of state-of-the-art AI 
models.

Recent accomplishments at the interface of  
physics inspired AI and supercomputing 
include the design of physics inspired 
AI architectures, training and optimization 
schemes that leverage thousands of 
GPUs~\cite{KHAN2020135628,Khan_HOM}. 
These AI surrogates have been used to process 
from seconds- to 
years-long datasets of gravitational wave data 
to demonstrate that AI  can be used to search for 
and find gravitational wave signals with 
an average false positive rate of one misclassification 
for every month of searched data~\cite{2021PhLB..81236029W,
huerta_nat_ast}. It has also 
been demonstrated 
that when these AI surrogates are optimized  
for accelerated inference with NVIDIA TensorRT, and the 
inference is distributed over the entire 
\href{https://www.alcf.anl.gov/alcf-resources/theta}{ThetaGPU supercomputer} at the Argonne Leadership 
Computing Facility, consisting of 160 NVIDIA A100 Tensor 
Core GPUs, gravitational wave data can be processed 
over $52,000\mathrm{X}$ faster than 
real-time assuming a two detector network comprising 
the twin Advanced LIGO detectors~\cite{Chaturvedi:2022suc}.

Since AI advances in gravitational wave astrophysics 
exhibit great promise~\cite{2023PhRvD.107b3021S,2018CQGra..35i5016R,Wei_quantized,Qiu:2022wub,2021PhRvD.103l3023L,2022PhRvD.105d3003S,2022PhRvD.105d3002S,Wei:2019zlc}, in this article we contribute to 
this line of research by designing AI 
architectures, training and optimization methods 
that incorporate physics and 
geometrical principles involved in the detection 
of gravitational waves. 
In practice, we have designed 
neural networks that capture both 
short- and long-range temporal dependencies of 
gravitational wave signals with hybrid dilated 
convolution networks. We also incorporate geometrical 
and spatial considerations of signal detection 
in terms of the location of gravitational wave 
detectors through graph neural networks. 
We show that this approach improves 
the sensitivity of AI ensembles for 
signal detection, while also reducing the number of 
false positives to 7 misclassification for 
every decade of searched data when using an ensemble of 
two AI models, and to 2 misclassification for every decade 
of searched data when using an ensemble of 4 AI models.
This is the first time AI methods achieve this 
level of accuracy over decade-long datasets.

We showcase this approach in the context of 
a network of three ground-based gravitational wave 
detectors encompassing the twin Advanced LIGO 
and Advanced Virgo detectors. 
We present results for an astrophysical population of 
quasi-circular, spinning, non-precessing 
binary black hole mergers. We show how to reduce 
time-to-solution by using distributed training on 
the \href{https://www.alcf.anl.gov/polaris}{Polaris supercomputer} 
at the Argonne Leadership Supercomputing Facility, 
in which we used 256 NVIDIA A100 GPUs to train AI models 
within 1.7 hours, while also ensuring 
optimal classification performance. We also demonstrate that 
our approach presents strong scaling up to 512 NVIDIA A100 
GPUs. We also used 128 NVIDIA A100 GPUs in Polaris 
and 128 nodes in the Theta supercomputer to process 
a decade's worth of simulated advanced gravitational wave 
data from a three detector networks within 3.5 hours, 
i.e., \(25,000\textrm{X}\) 
faster than real-time.

\section*{Results}

We present a novel approach 
that brings together 
physical and geometrical considerations in the 
design and training of AI models for 
gravitational wave detection. We use a 
hybrid dilated convolution network (HDCN) to 
capture long-range temporal dependencies that are 
crucial for signal prediction, and combine it 
with a graph neural 
network (GNN) to merge prediction 
embeddings from a three detection network---Advanced 
LIGO Livingston (L) and Hanford (H); and 
Advanced Virgo (V)---considering their spatial relationship.

We used the \texttt{IMRPhenomXPHM} 
waveform approximant~\cite{2021PhRvD.103j4056P} to 
model signals that include 
the higher-order wave modes
$(\ell, |m|)=\{(2, 2), (2, 1), (3, 3), 
(3, 2), (4, 4)\}$, and mode mixing effects 
in the \(\ell = 3, |m| = 2\) harmonics for 
quasi-circular, spinning, non-precessing binary black 
holes. Given the 
large amount of modeled waveforms needed to 
optimally sample this signal manifold, 
we introduce novel approaches to train these 
AI models at scale in modern computing 
environments. We showcase how to use these 
AI models to search for modeled waveforms 
in synthetic, recolored  noise; as well as in 
real gravitational wave data from the third 
Observing Run (O3) taken from the Gravitational 
Wave Open Science Center~\cite{KAGRA:2023pio}.
To the best of our knowledge, these are the first 
type of AI models designed to search for 
higher order gravitational wave modes in real 
gravitational wave data.

In the Methods section we describe in detail the 
datasets and modeled waveforms used for these 
studies, and approaches we have developed to train 
AI models at scale, and to enable hyper-efficient 
AI inference of large scale gravitational wave datasets 
in leadership class supercomputers. 

We present results that quantify the 
ability of our AI models to search for and find 
gravitational waves in a decade-long gravitational wave 
dataset in which we consider a three detector network 
comprising the twin Advanced LIGO detectors and the 
Advanced Virgo detector, assuming an astrophysical 
population of quasi-circular, spinning, 
non-precessing binary black hole mergers.

\subsection*{Quasi-circular, spinning, non-precessing  
binary black hole mergers}

We consider a binary black hole population 
described by 4 parameters, namely, 
\((m_1, m_2, s_1^z,s_2^z)\). 
The individual masses span the range 
\(  m_{\{1, 2\}} \in [3M_{\odot}, 50M_{\odot}]\), 
while their individual spins cover the range 
\(s_{\{1, 2\}}^z\in[-0.9,0.9]\). 
To instill 
the physics of these astrophysical sources 
into our AI surrogates, we have produced datasets of modeled 
waveforms using the \texttt{IMRPhenomXPHM} 
model~\cite{2021PhRvD.103j4056P}, which we used to incorporate 
higher order wave modes in the modeling of 
binary black hole mergers. We also used this model 
since it facilitates the production of 
modeled waveforms at scale in modern computing 
environments. Here, we used three independent, non-overlapping 
datasets: the training dataset has 
1.2 million waveforms, whereas the validation and test sets 
have 300,000 modeled waveforms. 

These waveforms 
are then curated to simulate 
a variety of astrophysically motivated scenarios. 
We do this by using tools 
provided by the open-source \texttt{PyCBC} 
library~\cite{pycbc_library} to uniformly 
sample the masses and individual spins of 
binary components, as well as the various angles 
that describe the sky location of these sources, 
except for the inclination angle for which 
we use a \(\sin\left(\mathrm{inclination}\right)\) 
distribution. 
For signal-to-noise 
ratios (SNRs) we use a uniform volume prior. 
This approach ensures that AI models are capable of 
correctly identifying gravitational waves that 
may be produced in a wide variety of astrophysical 
scenarios. Once our AI models are fully trained, we test them 
with new datasets that have not seen by these AI 
models during training. These new datasets contain 
either pure noise, or noise plus injected signals 
that simulate detection scenarios.  

\begin{figure}[htp]
    \centering
     \includegraphics[width=.9\textwidth]{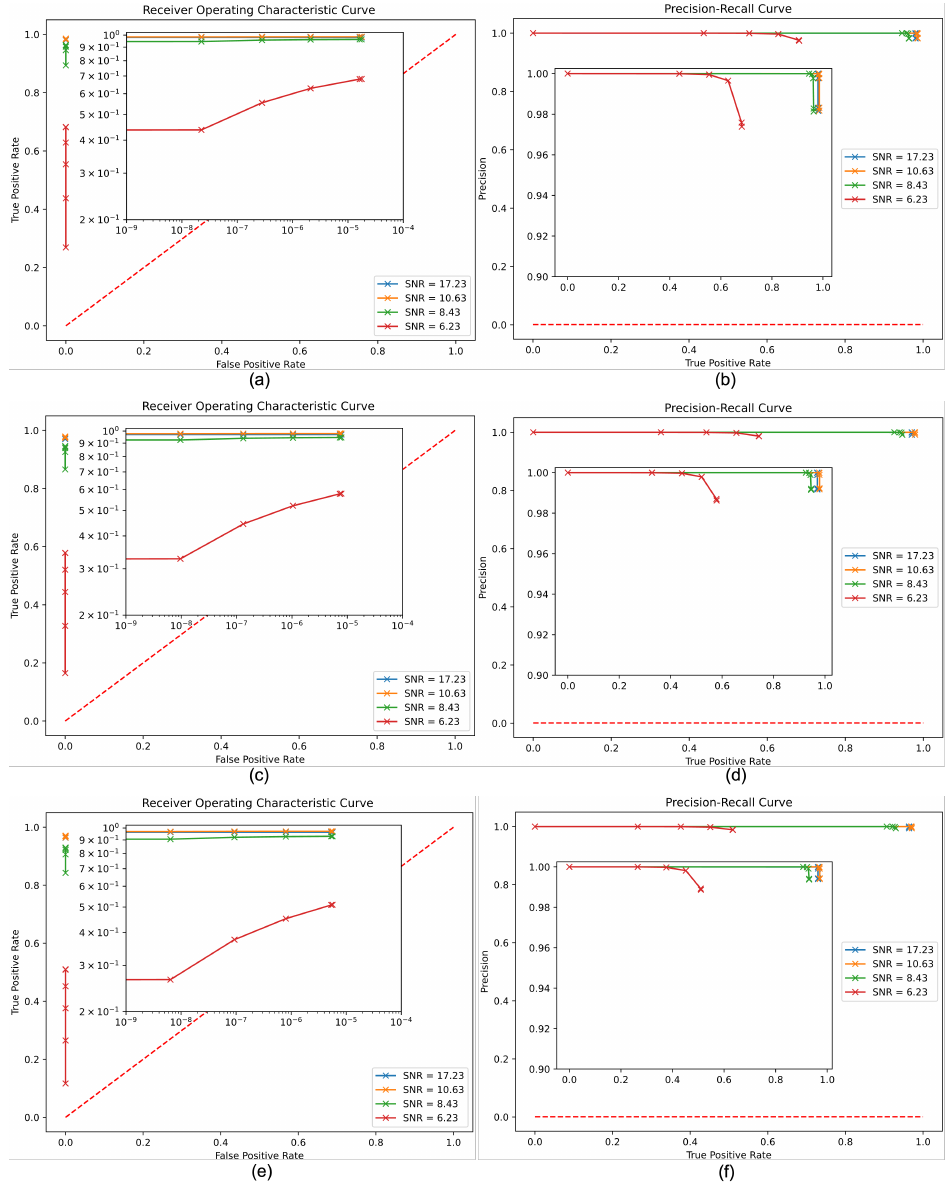}
    \caption{\textbf{Classification performance of AI ensembles to detect quasi-circular, spinning, non-precessing 
    black hole mergers.} Classification performance in terms of the 
    Receiver Operating Characteristic (ROC) curve, 
    and the Precision-Recall (PR) curve for ensembles that 
    include 2 AI models (top row), 3 AI models (center row), 
    and 4 AI models (bottom row). These results were produced 
    using a decade-long gravitational wave test set, 
    for the Advanced LIGO and Advanced Virgo three detector network, in which we injected 300,000 modeled binary black hole waveforms that cover a broad SNR range.}
    \label{fig:spin_2_n_4models}
\end{figure}

\noindent Figure~\ref{fig:spin_2_n_4models} 
presents two types of metrics we used to quantify the 
performance of our AI models for signal 
detection, i.e., 
the Receiver Operating Characteristic (ROC) curve 
and the Precision-Recall (PR) curve. Therein we 
present results for three cases, in which 
we construct ensembles that include, 2 AI models (top row), 3 
AI models (center row), and 4 AI models (bottom row), 
respectively. 

To create the ROC curve, we computed the 
true positive rate against the false positive rate as 
estimated from the output of our AI ensembles when 
we use them to process a decade-long dataset that 
describes a three detector network that comprises the 
Advanced LIGO and 
Advanced Virgo detectors. As mentioned before, 
the test set includes 300,000 modeled waveforms to 
quantify the ability of AI ensembles to correctly identify 
modeled waveforms, and discard other noise anomalies. To 
produce the PR curve, we consider that $\textrm{PR}= \textrm{TP}/\left(\textrm{TP}+\textrm{FP}\right)$, 
where $\textrm{TP}$ and $\textrm{FP}$ stand for 
True Positives and False Positives, respectively. Key results we extract from these studies include the following:

\begin{itemize}[nosep]
    \item Figure~\ref{fig:spin_2_n_4models} shows 
    that AI's precision to detect gravitational waves 
    improves as we increase the number of 
    AI models in the ensemble (right column). At 
    the same time, there is a marginal reduction in the 
    true positive rate, in particular for  
    low SNR signals (both left and right column).
    \item We provide key figures of merit in Table~\ref{tab:fps_dec_spin}. Therein we show 
    that a 2 AI model 
    ensemble reports 7 false positives per decade, 
    a 3 AI model ensemble reports 3 false positives 
    per decade, while a 4 AI model ensemble produces 
    2 misclassifications per decade of searched data. 
    We also present the Area Under the Curve (AUC) for 
    the ROC and PR curves for gravitational wave 
    signals in our 300,000 test set with 
    cumulative 
    three detector network $\textrm{SNR}_1=8.4$ 
    and $\textrm{SNR}_2=10.6$. 
    We selected these cumulative 
    three detector network SNR values since they provide a fair 
    representation of the bulk of events reported in 
    the Gravitational Wave Open Science 
    Center~\cite{Vallisneri:2014vxa}. These results 
    show that as we consider larger AI ensembles 
    there is marginal decrease in the ROC AUC and PR AUC 
    for low SNR signals, while signals with $\textrm{SNR}_2=10.6$ 
    have a negligible change. 
\end{itemize}

\begin{table}
\caption{Classification performance of AI ensembles 
that include 2, 3 or 4 AI models (left column). 
Here we consider a 
population of quasi-circular, spinning, non-precessing binary black hole 
mergers. FP presents the number of false positives 
reported by the AI ensembles upon 
processing a decade-long gravitational wave test set 
that describes a three  
detector network comprising the Advanced LIGO 
and Advanced Virgo detectors, and in which we injected 300,000 
modeled waveforms that cover a broad range of 
signal-to-noise ratios (SNRs). We also present additional 
figures of merit, namely, the Area Under the Curve (AUC) for 
the Receiver Operating Characteristic (ROC) curve, 
and the Precision-Recall (PR) curve for two 
cumulative three detector network SNR values 
that represent the bulk of events detected thus far 
by Advanced LIGO and Advanced Virgo, i.e., $\textrm{SNR}_1=8.4$, 
and $\textrm{SNR}_2=10.6$---see also Figure~\ref{fig:spin_2_n_4models}.}
\begin{indented}
\lineup
\item[]\begin{tabular}{@{}*{6}{c}}
\br                       
    \# of models &
    FP &
    \centre{2}{ROC AUC} &
    \centre{2}{PR AUC} \\
      &  & {$\textrm{SNR}_1$ } & {$\textrm{SNR}_2$} & {$\textrm{SNR}_1$} & {$\textrm{SNR}_2$} \\
    2 & 7  & 0.9818 & 0.9929  & 0.9636 & 0.9857 \\
    3 & 3  & 0.9730 & 0.9894  & 0.9460 & 0.9789 \\
    4 & 2  & 0.9649 & 0.9855  & 0.9297 & 0.9710 \\
    \br
  \end{tabular}
\label{tab:fps_dec_spin}   
\end{indented}
\end{table}

\noindent This work significantly outperforms recent 
results in the literature~\cite{huerta_nat_ast,Chaturvedi:2022suc}, 
in which AI models were used to process a 5 year-long dataset. 
Here we significantly improve the classification 
performance of those AI models, both in terms of 
improving the true positive rate for signals 
across SNRs, and with a very significant improvement 
for low and moderate SNR signals (compare these findings to 
those in Figure 4 in~\cite{huerta_nat_ast}). We also 
reduce the number of misclassifications from 1 per month to 
2 per decade. This is the first time AI reaches this 
level of accuracy to correctly tell apart gravitational 
waves from other noise anomalies. 

In brief, AI ensembles trained with 
modeled waveforms that describe quasi-circular, spinning, 
non-precessing binary black hole mergers 
provide optimal results. This is because 
these AI models are exposed to signals with more 
complex morphology and time-evolution, and thus 
are better equipped at telling apart true signals 
from other noise anomalies. Again, if compare 
these results to those presented in Refs.~\cite{huerta_nat_ast,Chaturvedi:2022suc}, we 
notice that the AI models and computational methods we 
introduce in this paper significantly improve the capabilities 
of AI for signal detection by 
enhancing its precision and accuracy, and by 
significantly reducing the 
number of false positives over long datasets. 

\paragraph{Results with synthetic noise: trends and patterns.} 
In the results above, we notice that there is a trade-off in 
true positive rate and false positive rate that is driven by 
the size of the ensembles. As we increase the number of AI models, 
the precision increases, i.e., we find less false positives. 
At the same time, the true positive rate also marginally decreases. Another factor at play here is that we are combining several 
AI models that share the same architecture, and whose initial 
weights have been randomly initialized for training and 
optimization. These three factors affect the ROC AUC and PR AUC. 
Note, though, that these differences are \(\leq3\%\), 
which is well within the range of fluctuations in performance 
due to training and optimization.

\subsection*{AI ensembles on O3b data}

In the results presented above, we considered a detector 
network whose individual components have comparable 
sensitivity. We now explore the application of our 
AI ensembles for O3b gravitational wave data in which 
the H, L and V detectors reported a binary neutron 
star inspiral range of 
135 Mpc, 115 Mpc, and 50 Mpc, 
respectively~\cite{LIGOScientific:2021djp}. These 
disparate sensitivities mean that our AI models 
will now have to deal with two input channels that provide 
comparable outputs, HL, while another channel, V, will 
provide noisier outputs. Since our AI architecture 
aggregates the output predictions from each 
H, L and V detector, this means that our AI models may get 
confused when H\&L predictions agree on the existence of a potential 
event, while V may flag it inconsistently 
as a positive or negative sample. Notwithstanding 
these significant changes in 
the original scope and design of our AI models, we have 
fine-tuned our AI models to search for HLV events in O3b data.

In practice, we selected O3b data covering the GPS times 
1264291218 through to 1267056018. For training 
purposes, we extracted three separate L, H, and 
V data segments, which are 4,096 seconds long, 
and have the following GPS start times: 1264685056, 
1265528832 and 1266200576. We estimated training 
PSDs for L, H, and V using these three 4,096 second 
long segments. To produce AI ensembles with real data, 
we considered 
the training dataset of 1.2 million higher order 
wave modes that describe quasi-circular, spinning, 
non-precessing binary black hole mergers.

\begin{figure*}[!htpt]
    \centering
    \includegraphics[width=\textwidth]{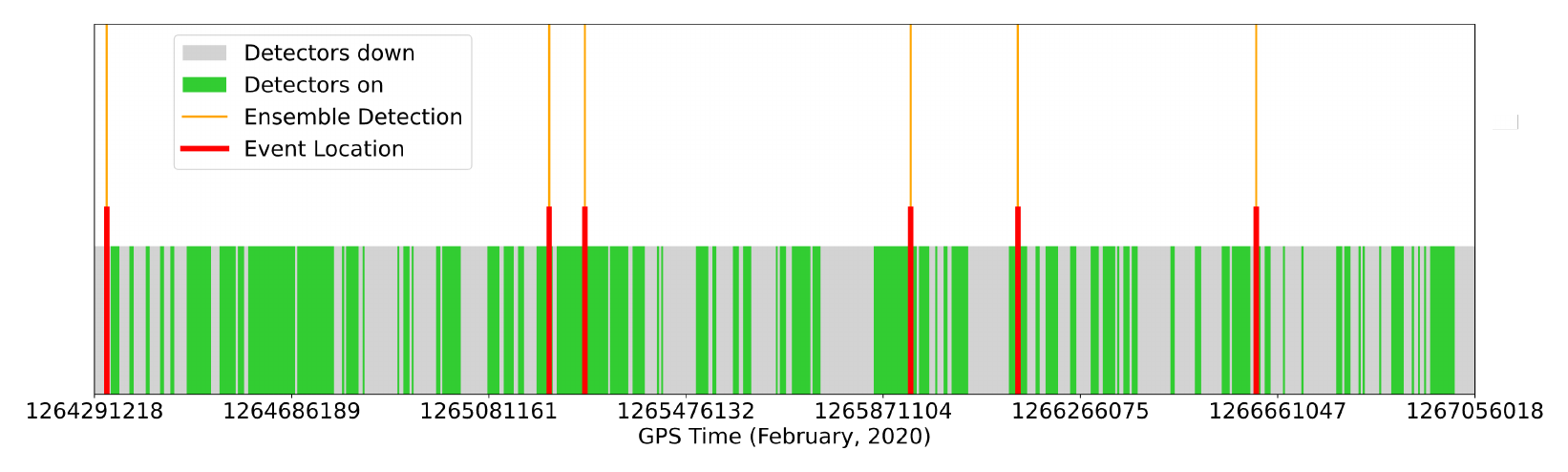}
    \caption{\textbf{Classification performance of AI ensemble throughout February 2020.} Our 6 AI model ensemble was able to identify 6 HLV events in February 2020 with no false positives. The 
    detected events from left to right are: GW200129\_065458\_1264314069,
GW200208\_130117\_1265200048,
GW200209\_085452\_1265271663,
GW200216\_220804\_1265924055,
GW200219\_094415\_1266138626, and 
GW200224\_222234\_1266616125.}
    \label{fig:real_data}
\end{figure*}

We then fine-tuned a suite of 20 AI models with these 
real datasets. Each of these models was fine-tuned 
within 320 minutes using 48 NVIDIA A100 
GPUs in the Polaris supercomputer. We selected the top 
6 AI classifiers and used them to process data from 
February 2020. Our findings indicate that this AI ensemble 
can identify the following 6 gravitational wave events: 
GW200129\_065458\_1264314069,
GW200208\_130117\_1265200048,
GW200209\_085452\_1265271663,
GW200216\_220804\_1265924055, \hfill \break 
GW200219\_094415\_1266138626, and 
GW200224\_222234\_1266616125. Our AI ensemble also 
reported no false 
positives, as shown in Figure~\ref{fig:real_data}.

\subsection*{Quantitative comparison to other signal detection pipelines with real data}

We can readily compare the output of our AI ensemble with 
other established pipelines for gravitational wave 
detection for the O3b data we analyzed above. These 
results are summarized in Table~\ref{tab:comp_AI_Oth}. 
It is worth pointing out that 
while the template-matching and 
template-agnostic pipelines listed in 
Table~\ref{tab:comp_AI_Oth} benefit from additional 
information such as data calibration, data quality 
information, and 
data cleaning, our AI ensembles are trained, 
validated and tested using available data 
at the Gravitational 
Wave Open Science Center~\cite{KAGRA:2023pio}. Thus, 
we have carried out an additional comparison 
between \texttt{PyCBC} and our AI ensemble using 
open source data as is. 

\begin{table}
\caption{HLV events detected by six different detection pipelines using data from the O3b run throughout February 2020. Note that single detection triggers are not 
included here.}
 \begin{indented}
\lineup
\item[]\begin{tabular}{@{}*{7}{c}}
\br   
 HLV Events & AI  & \texttt{cWB} & \texttt{GstLAL} & 
 \texttt{MBTA} & \texttt{PyCBC-broad} & \texttt{PyCBC-BBH}  \\
 \hline
 GW200129\_065458\_1264314069   & \checkmark & & \checkmark &  &\checkmark  & \checkmark \\
 GW200208\_130117\_1265200048   & \checkmark &   &\checkmark& \checkmark & \checkmark& \checkmark\\
 GW200209\_085452\_1265271663   & \checkmark &   &\checkmark& \checkmark & \checkmark& \checkmark\\
 GW200216\_220804\_1265924055   & \checkmark &   &\checkmark& \checkmark & \checkmark& \checkmark\\
 GW200219\_094415\_1266138626   & \checkmark & \checkmark  &\checkmark& \checkmark & \checkmark& \checkmark\\
 GW200224\_222234\_1266616125   & \checkmark & \checkmark  &\checkmark& \checkmark & \checkmark& \checkmark\\
 \br
\end{tabular}
\end{indented}
\label{tab:comp_AI_Oth}
\end{table}

\begin{table}
\caption{Signal detection comparison between our 
AI ensemble and \texttt{PyCBC} analyses for 10 
events reported in Ref.~\cite{LIGOScientific:2021djp} 
for O3b data covering the GPS times 
1264291218 through to 1267056018. The data 
used for this analysis is open source data 
available at the Gravitational 
Wave Open Science Center~\cite{KAGRA:2023pio}.}
\begin{indented}
\lineup
\item[]\begin{tabular}{@{}*{5}{c}}
\br                              
 HLV Events & AI & \# False Positives & \texttt{PyCBC} & \# False Positives \\
 \hline
 GW200129\_065458\_1264314069   & \checkmark & 0  & \checkmark & 307  \\
 GW200202\_154313\_1264691364   & & 0  & & 239  \\
 GW200208\_130117\_1265200048   & \checkmark & 0  & \checkmark & 284  \\
 GW200208\_222617\_1265233948   & & 0  & & 323  \\
 GW200209\_085452\_1265271663   & \checkmark & 0  & & 291  \\
 GW200210\_092254\_1265359745   & & 0  & & 290  \\
 GW200216\_220804\_1265924055   & \checkmark & 0  & & 284  \\
 GW200219\_094415\_1266138626   & \checkmark & 0  & & 219  \\
 GW200220\_061928\_1266212739   & & 0  & & 231  \\
 GW200224\_222234\_1266616125   & \checkmark & 0  & \checkmark & 273  \\
 \hline
\end{tabular}
\end{indented}
\label{tab:comp_AI_pycbc}
\end{table}

\noindent To carry out this analysis, we analyzed 4096s of HLV data around each of the 10 
events reported in Ref.~\cite{LIGOScientific:2021djp} 
for O3b data covering the GPS times 
1264291218 through to 1267056018. Using 
\texttt{PyCBC} version 
32498060a5ef01fad606fe0a3e28b1a632adff30 
(Nov 6, 2023), we computed the 
matched-filtering integral from 20Hz with the 
data and templates being sampled at 2048Hz. 
We used \texttt{IMRPhenomD} waveform templates~\cite{Khan:2015jqa} 
with a bank that covers the black hole mass range  
\(  m_{\{1, 2\}} \in [3M_{\odot}, 50M_{\odot}]\), 
and aligned individual spins 
\(s_{\{1, 2\}}^z\in[-0.9,0.9]\) with a minimal match of 
97\%. The search estimates the noise power spectral 
density over 512 second chunks of data, and takes the 
median over all chunks. Other technical details, 
including the ranking statistic used and methodology 
to compute the significance of triggers are as used in 
LVK analyses~\cite{LIGOScientific:2021djp}. Note that 
our AI ensemble is designed to consider aligned 
individual spins \textit{and higher order modes}, 
which are not considered in this \texttt{PyCBC} 
since it would considerably increase its 
computational cost.  

The \texttt{PyCBC} runs for each of the 4096s long 
segments include filtering, coincidence finding, 
ranking of events, and a small number of injections 
to determine the horizon range of the search. Each 
individual analysis was completed in 4.5 hours using 
4 CPUs with 8GB RAM. AI inference on each 
of these segments was completed in 20.95s with one 
NVIDIA A100 GPU. We summarize the findings 
of this analysis in Table~\ref{tab:comp_AI_pycbc}. 
Therein we notice that our AI ensemble can identify 
6 HLV gravitational wave events with 
no misclassifications. \texttt{PyCBC} can identify 
3 gravitational wave events, and each has a sizeable list of 
false positives, which are directly provided by \texttt{PyCBC}, 
and correspond to coincident triggers present in all three 
detectors, HLV, that have a minimum 
individual detector SNR of 5. We took a closer look 
at the 4 events that our AI ensemble did not find, 
and discovered that:

\begin{itemize}[nosep]
    \item GW200202\_154313\_1264691364. According 
    to Ref.~\cite{LIGOScientific:2021djp}, this event had 
    individual detector SNR of 4.6, 10, and 2.4 for H, L, and V, 
    respectively, for the GstLAL pipeline~\cite{CANNON2021100680}. Since our AI model 
    aggregates the predictions from the three detector outputs, 
    this event went unnoticed for both the  
    H and V nodes, and thus marked as noise. 
    \item GW200208\_222617\_1265233948. As before, 
    according to Ref.~\cite{LIGOScientific:2021djp}, this event had 
    individual detector SNR of 5.6, 5.7, and 2.1 for H, L, and V, 
    respectively, for the GstLAL pipeline~\cite{CANNON2021100680}. This event 
    went unnoticed for V and thus marked as noise.
    \item GW200210\_092254\_1265359745. This is a 
    neutron star-black hole system, outside of the 
    parameter space that we considered in this study. 
    \item GW200220\_061928\_1266212739. The masses of 
    this event are out of the range we considered to 
    train our AI models. 
\end{itemize}

\noindent In brief, this analysis indicates that 
our AI ensembles may greatly benefit from using 
additional information to further increase 
their detection capabilities, i.e., data quality 
information, data cleaning, and other 
methodologies used in established 
gravitational wave detection pipelines. We also 
learn that AI methods already provide the means to 
search for gravitational wave signals with 
very complex morphology at minimal computational cost.

\section*{Methods}

Here we describe in detail the datasets, 
AI architectures, training and optimization 
schemes used to create our AI models.

\paragraph{Datasets.} We used the 
\texttt{IMRPhenomXPHM} waveform 
model~\cite{2021PhRvD.103j4056P} to create 
datasets of modeled waveforms, sampled at 
4096 Hz, 
that describe an astrophysical 
population of binary black holes that spiral 
into each other following a series of 
quasi-circular orbits. Their binary components 
span the masses 
\(m_{\{1, 2\}}\in [3M_{\odot}, 50M_{\odot}] \), 
and individual spins \(s_{\{1,2\}}\in[-0.9, 0.9]\). 
We generated 1.8 million waveforms 
by uniformly sampling this \((m_1, m_2, s_1^z,s_2^z)\) 
parameter space. Furthermore, we incorporated 
information about the sky location and 
detector location into the modeled waveforms by 
sampling right ascension and declination, 
orbital inclination, coalescence phase, and 
waveform polarization. Right ascension and declination 
are sampled uniformly on a solid angle of a sphere. 
The polar angle varies from $\pi/2$ (north pole) to 
$-\pi/2$ (south pole). The orbital inclination is 
sampled using a $\sin(\textrm{inclination})$-distribution. 
The coalescence phase and waveform polarization are 
both uniformly sampled covering the range $[0,2\pi)$. 
We have densely sampled these parameters to expose our 
AI surrogates to a broad range of plausible 
detection scenarios. With this approach, our 
AI surrogates learn the interplay between 
these parameters, and the slightly different times 
at which the merger event (waveform amplitude peak) is 
recorded by our AI surrogates in their three input 
data channels, representing the Advanced LIGO and 
Advanced Virgo detectors.

For the construction of AI ensembles using 
real data, we also consider modeled waveforms 
that describe higher order wave modes for 
quasi-circular, spinning, non-precessing binary black hole mergers.

Out of the 1.8 million waveforms in our dataset, 
we have created three independent, non-overlapping 
datasets. The training set has 1.2 million waveforms, 
whereas the validation and test sets have 
300,000 modeled waveforms each. In practice, 
we produced \texttt{HDF5} files each containing 2,000 modeled 
waveforms that uniformly sampled the parameter space with 
random variable samples. We then split these \texttt{HDF5} 
files into training, validation and test sets, ensuring 
that they provide a fair coverage of the parameter space. 
We follow best machine learning practices and 
ensure that the recolored noise and signals 
used to produce the training, validation and test sets 
are different to each other.

\paragraph{Power Spectral Density for Advanced LIGO and Advanced Virgo.} 
To model the sensitivity of Advanced 
LIGO and Advanced Virgo detectors, we rely on 
Power Spectral Density (PSD) curves. Specifically, 
we use PSD curves from Advanced LIGO Sensitivity 
(190 Mpc) based on the fourth observing run (O4, aligo\_O4high.txt) 
for the Hanford and Livingston detectors. For Advanced 
Virgo detector, we use PSD data from the fifth 
observing run (O5, avirgo\_O5low\_NEW.txt), 
which has a low noise and high 
range limit target sensitivity~\cite{Abbott2020,pdsnoise}. We selected 
these PSDs to study detection scenarios in 
which the LIGO and Virgo detectors have 
similar sensitivities, which may be attained 
by the end of O5. As we also show below, if 
one considers O3b PSDs, then the Advanced LIGO 
and Advanced Virgo 
detectors have rather disparate sensitivities 
and the AI models get confused when 
a noise trigger is found in LIGO data streams 
(L \& H), whereas the same noise trigger in 
Virgo data is inconsistently labeled by the AI 
models, especially 
for moderate SNR signals, given the current  
noisier nature of Advanced Virgo datasets.

In the context of real gravitational 
wave data, we estimated training PSDs for the H, L and V 
detectors using three 4,096 seconds long segments 
with GPS start times 1264685056, 
1265528832 and 1266200576. These segments do not contain 
any known gravitational wave signals. Following 
best machine learning practices, the training, validation 
and test sets are independent, i.e., there is no 
overlap among them in terms of the simulated signals and real 
gravitational wave noise used in each of these datasets. 

\paragraph{Data curation.} Modeled waveforms 
were whitened with the aforementioned PSDs. We also 
produced synthetic noise, and colored it with the 
PSDs representing each of the detectors. To address the challenge posed by highly imbalanced data in real detection, our dataset uses 70\% negative samples (only noise) 
and 30\% positive samples (noise and signals). 

For generating negative samples in our model, 
we randomly select 1s-long segments of pure noise
from the generated synthetic noise. These negative samples 
are labeled as pure 0s. On the other hand, for generating 
positive samples, we consider the whitened signal to the
0.5s before the merger and the ringdown part. Thereafter, we 
linearly mixed whitened waveforms and whitened noise so as to 
describe a broad range of astrophysical scenarios. The label 
for positive samples will be set to 1 only for the 0.5s before 
the merger, while the remaining part will be labeled as 0, 
since we are mostly interested in the merger portion 
of the signals where the AI models have the sharpest response. 

For data curation in the 
context of gravitational wave data, we follow the same 
exact procedure with the exception that now 
we use real H, L and V data to noise-contaminate modeled 
waveforms, and we whitened both signals and noise using the 
PSDs that we estimated with real data.

\begin{figure}[!htbp]
    \centering
    \includegraphics[width=0.9\textwidth]{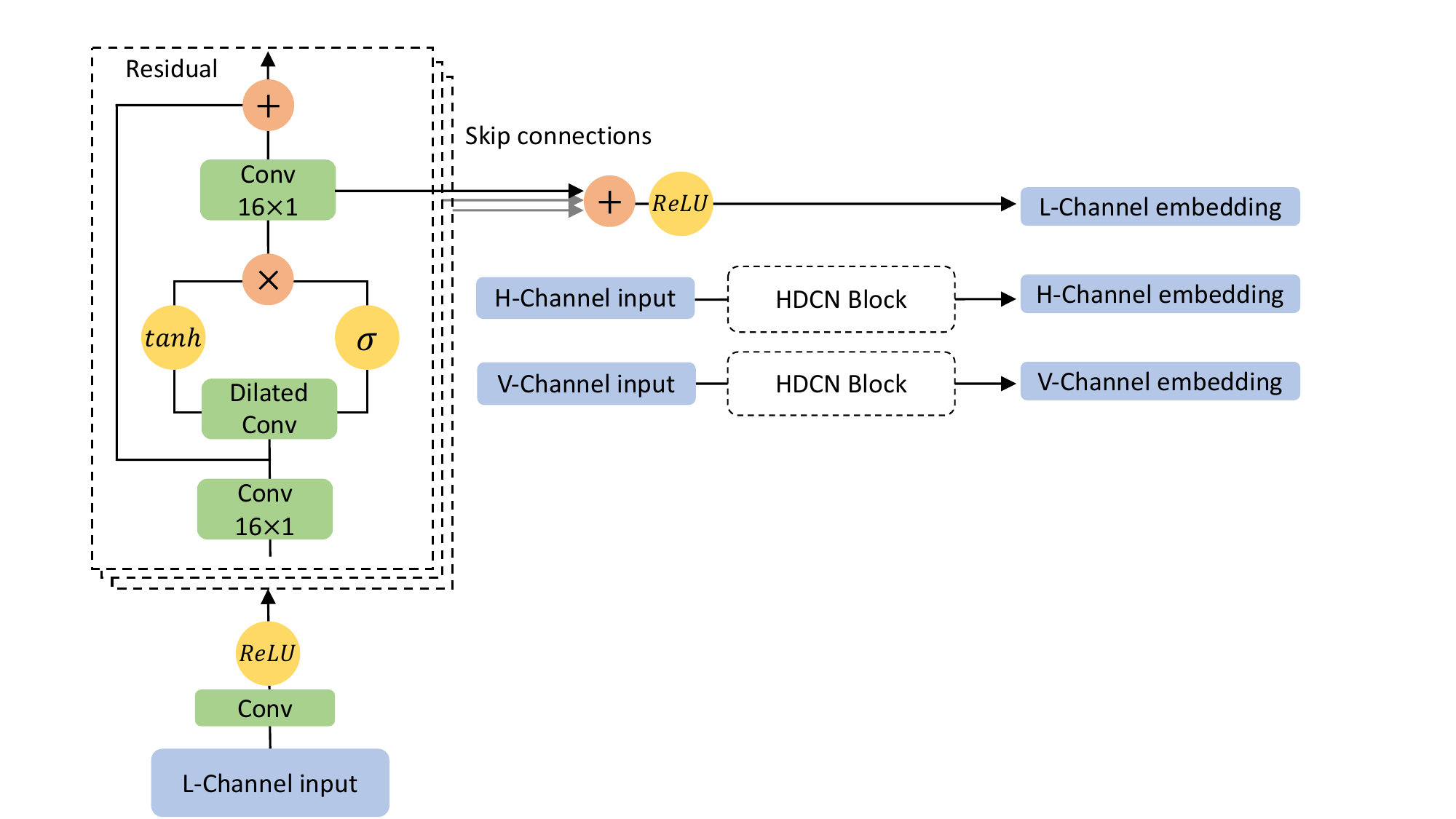}
    \caption{\textbf{Architecture of hybrid dilated 
    convolution network (HDCN).} Each detector input will 
    be processed by an individual HDCN, which consists of the N 
    blocks of pre-processing time-distributed convolution 
    layer, dilated convolution layer, and the post-processing 
    time-distributed convolution layer. N depends on the 
    receptive field desired. Here we use N=11 to cover a half second receptive field. Each block has residual structure 
    and skip connection to the output embedding.}
    \label{fig:HDCN}
\end{figure}

\paragraph{AI model architecture.} Our 
neural network model has two building blocks: 
the HDCN block, and the GNN block, which generates 
the final prediction with an output layer. Our HDCN block is inspired by WaveNet~\cite{2016wavenet}, 
a deep neural network for audio generation tasks, 
such as human speech or music. It uses dilated causal 
convolutions to model the temporal dependencies of audio 
signals, allowing it to generate high-fidelity audio with 
long-range structure. We believe we can also use it for our 
high sample rate time-series data, as the dilated 
convolution layers enable the network to capture larger 
receptive fields using fewer parameters. Moreover, its 
block-based structure enables the model to respond to a 
diverse range of frequency bands.

\begin{figure}[!htbp]
    \centering
    \includegraphics[width=\textwidth]{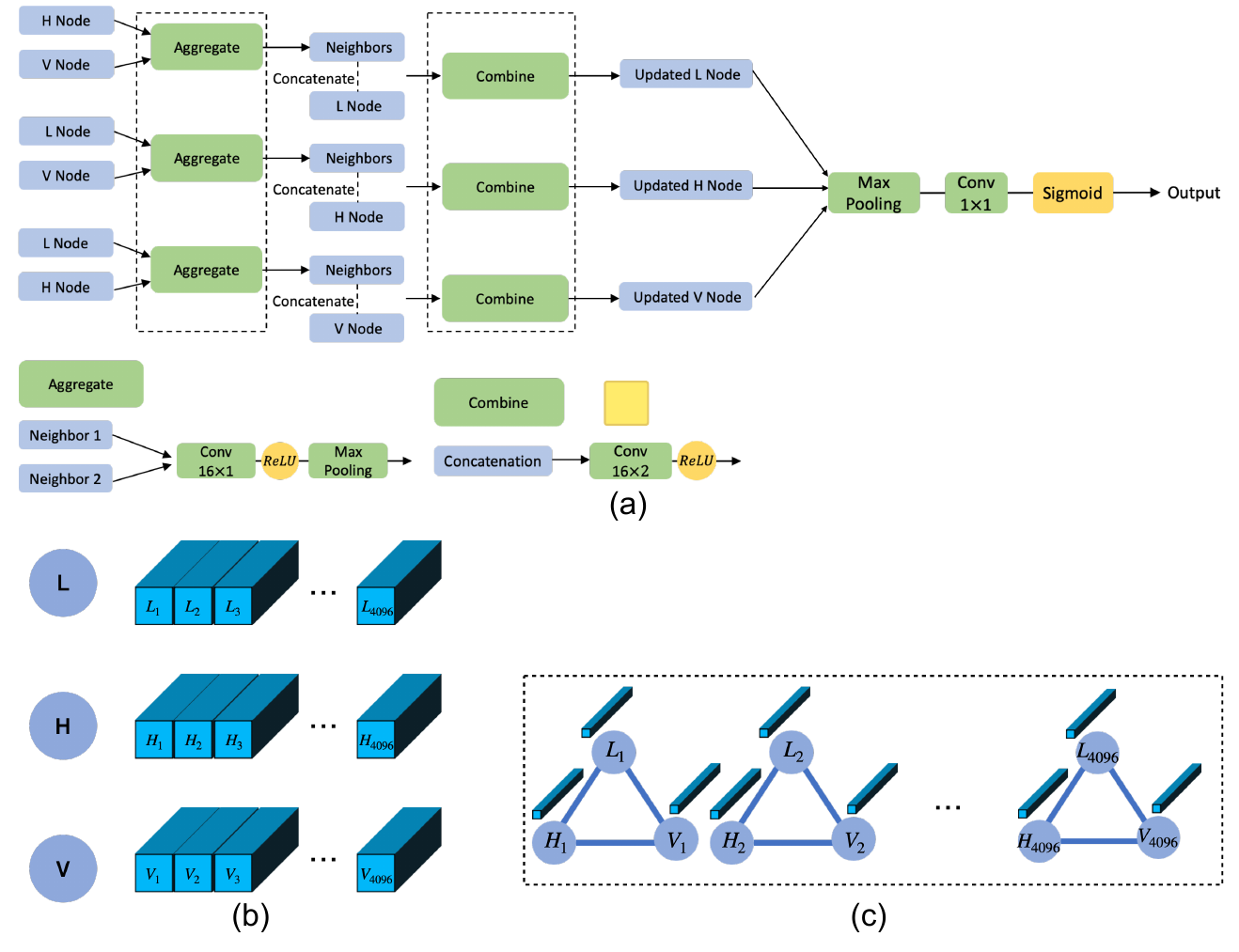}
    \caption{\textbf{Spatiotemporal-graph AI model structure.} 
    (a) Graph neural network (GNN) structure,
    (b) 3 prediction embeddings given by the hybrid dilated 
    convolution network (HDCN) block,
    (c) Geometric visualization for prediction embeddings 
    of the three time series produced by the Advanced LIGO and 
   Advanced Virgo detectors.}
    \label{fig:GNN}
\end{figure}

We present the hybrid dilated convolution structure in 
Figure~\ref{fig:HDCN}. It consists of pre-processing and 
post-processing time-distributed convolution layers and 
dilated convolution layers. With time-distributed convolution 
layers, HDCN is able to enlarge the expressiveness of 
the embedding. By stacking dilated convolution layers 
with different dilation rates and gated activation unit, HDCN 
will get an exponential increase in the size of the receptive 
field with respect to the number of layers. It will be crucial 
when considering long-range temporal dependencies in our 
signals. Finally, residual structure and skip connections 
are used to combine both long-range and short-range 
dependencies. Every detector strain data will be processed 
by an i.i.d. HDCN block to output the prediction 
embeddings since we want to preserve each detector's own 
information for the following GNN block.

On the other hand, there are various GNN structures 
available~\cite{bruna2013spectral, gilmer2017neural, NIPS2017_5dd9db5e, velickovic2017graph, xu2018powerful}, 
and we decided to 
use the Message Passing Neural Network (MPNN) framework with 
max pooling as the permutation invariant operator. We believe 
this can help capture the most important features in the 
embeddings and discard the less significant ones. Additionally, 
max pooling can help make the model more robust to small 
perturbations or noise in the input data by focusing on the 
strongest features, i.e., on the merger event.

Following the HDCN blocks, we obtain three 
prediction embeddings 
from the three detector channels. Since all three embeddings 
should represent the same true signal generated by the binary 
black hole merger, we need to combine them to produce the 
final prediction output. While a naive approach is to 
concatenate the three channels and pass them through a 
multi-layer perceptron (MLP)~\cite{2021PhLB..81236029W}, this 
method is not entirely optimal. If we change the order of the 
channels, the MLP output will differ even though the channels 
still represent the same signal. Therefore, we need a more 
robust approach to combine the embeddings. We use GNN to 
preserve the permutation invariance between the three detector 
embeddings and obtain a graph-level embedding 
as the combined output.

The GNN structure is shown in Figure~\ref{fig:GNN} (a). 
For each target node, the GNN block applies an Aggregate
step and a Combine step. During the Aggregate step, 
the GNN processes each neighbor embedding using a convolution 
layer and then uses 
max pooling to output the aggregated neighbor message, which 
is permutation invariant. In the Combine step, the GNN 
concatenates the aggregated neighbor message with the target 
node embedding and passes the result to another convolution 
layer to update the target node embedding. The Aggregate and 
Combine layer is shared among the 3 nodes. Next, we use max 
pooling to combine the 3 node embeddings and generate a 
graph-level embedding. Finally, we apply an MLP with sigmoid 
activation to the graph-level embedding to generate 
element-wise predictions.

In the GNN block, we 
represent each 
time step as a three-node graph and use the GNN to combine the 
three embeddings of each time step. Since each time step 
already encompasses a long receptive field after HDCN, 
see Figure~\ref{fig:GNN} (b), this 
method is sufficient to generate accurate predictions at 
each time step. We can easily incorporate multiple-step 
correlation by increasing the kernel size of the convolution 
layer accordingly. Then, we apply a general GNN to process 
the node embeddings, 
as it captures the relationship between the embeddings, i.e., 
the locations of the detectors. Since the detector locations remain constant over time, we use the same GNN function for all time steps. We can relax this assumption when we consider signals whose detector antenna patterns evolve over time. 
This approach has the added benefit of 
preserving the strong correlation between the embeddings, 
which is due to their highly overlapping receptive fields. 
Consequently, we can maintain this correlation in our final 
time-series prediction output. We can consider each time step 
as a subgraph in a larger graph, as shown in 
Figure~\ref{fig:GNN} (c), allowing us to share the 
aggregation and combination functions across all time steps. 
In other words, we propose a time-independent GNN to process 
the time-series embeddings.

\paragraph{AI ensembles.} Our ensemble method employs a 
cascading strategy where detection is considered positive, i.e., a real signal or an injected signal, only if there is unanimous agreement among all models. Conversely, if even a single model identifies the detection as negative, i.e., pure noise, the ensemble treats the entire detection as negative. This stringent criterion for positive detection is designed to minimize FPs, ensuring that only the most reliable detections are classified as positive. By requiring consensus, this approach aims to enhance the specificity of our system, effectively reducing the likelihood of false alarms without significantly compromising the sensitivity. This method leverages the strengths of multiple models, filtering out less confident predictions, thereby achieving a balance between sensitivity and specificity that is crucial for maintaining high performance in our application.
In practice, when we consider AI models trained, valiated, and 
tested with synthetic noise, we find that AI ensembles with one, two, three, and four AI models produce 599, 7, 3, and 2 FPs when tested over a decade long dataset.

\paragraph{Detection strategy.} For the determination of the threshold used in calculating FPs, we strategically select the threshold that corresponds to the point on the ROC curve nearest to the top left corner, i.e., closest to the point (0,1), which is used to determine the width and height of the  \texttt{find\_peaks} algorithm, where we choose 0.5 seconds 
for the width and a threshold of 0.999999 for 
the height of the peak. This decision is grounded in the empirical observations from our validation dataset, where our model demonstrates a strong capability in identifying positive cases even at lower SNR levels, yet accompanied by a considerable number of FPs. Given this performance characteristic and the anticipation of encountering more pronounced data imbalance in real scenarios---predominantly featuring negative or pure-noise samples---we opted for a threshold embodying near-absolute confidence. This stringent threshold criterion aims to substantially minimize false positives, enhancing the precision of our model in real-world applications where the rarity of true positives necessitates a highly cautious approach to declaring positive detections. This method of threshold selection, while potentially increasing the model's specificity, is a deliberate choice to ensure the reliability and applicability of our model's outputs, especially in contexts where the cost of false positives is high and the occurrence of true positives is exceptionally rare.

\paragraph{Distributed training at scale.}
We use data parallelism, but not model parallelism, 
for both training and inference since our model can be fit into a single A100/V100 GPU. In the training stage, data parallelism can help accelerate training time. By distributing the workload across several GPUs, data parallelism significantly reduces the time required for training. This acceleration is crucial for our large datasets, where single-GPU training would be prohibitively slow.
Also, the framework for data parallelism is inherently scalable. It can accommodate additional GPUs without major changes to the training algorithm, allowing for linear or near-linear speed-ups with increased hardware resources.

\begin{figure}[htp]
    \centering
    \includegraphics[width=0.8\textwidth]{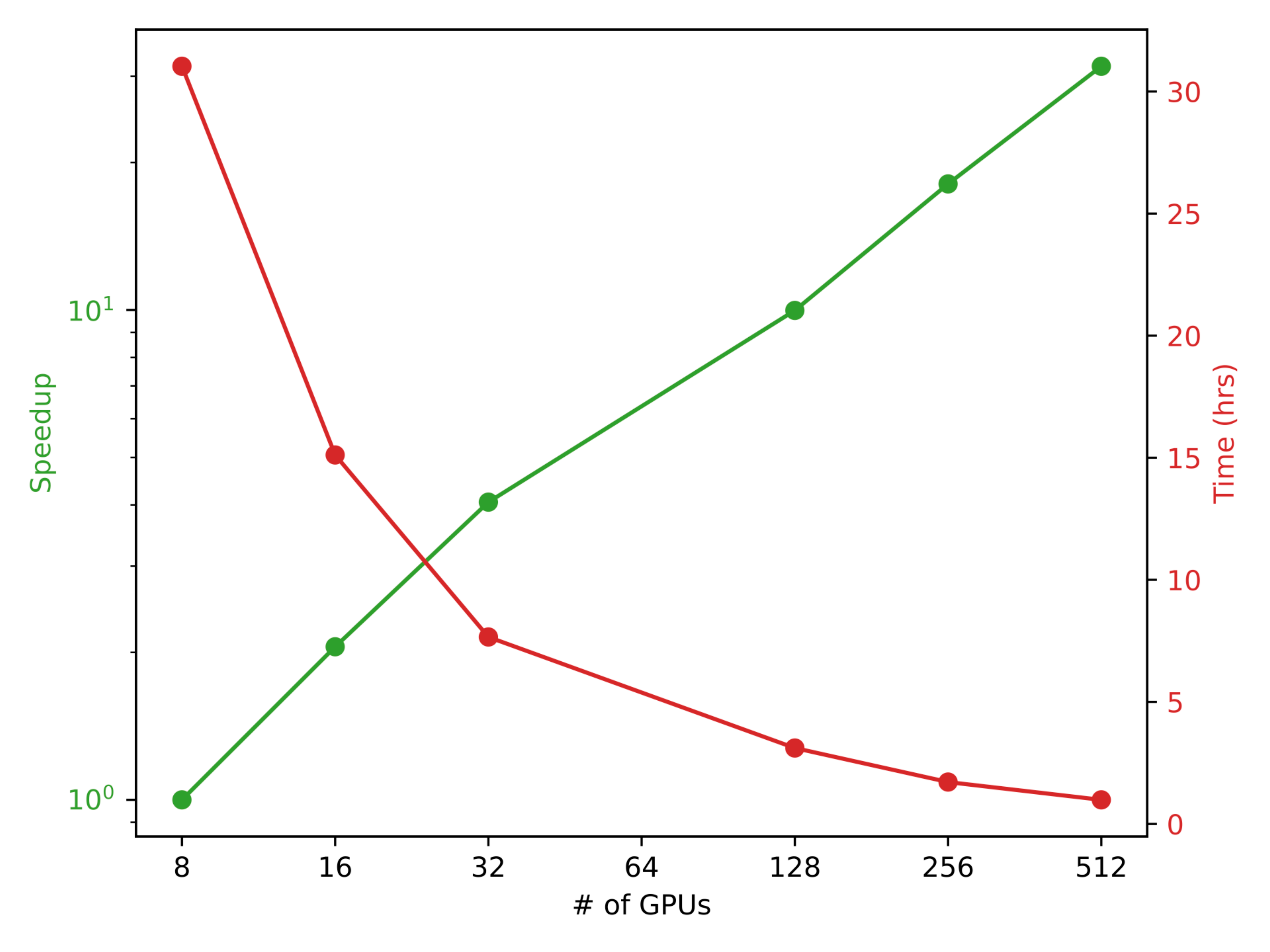}
    \caption{\textbf{Distributed training in the Polaris supercomputer.} We developed novel methods to 
    complete the training of spatiotemporal-graph 
    AI models within 1.7 hours using 256 A100 NVIDIA GPUs. 
    Our optimization method exhibits strong scaling up to 512 NVIDIA A100 GPUs.}
    \label{fig:scaling}
\end{figure}

We conducted scaling tests in the Polaris 
supercomputer at the Argonne Leadership Computing 
Facility. In Figure~\ref{fig:scaling} we present 
our scaling results. The AI models presented in 
this article were trained within 1.7 hrs using 
256 A100 GPUs. All these models have 
optimal classification accuracy. We also conducted 
scaling studies using up to 512 A100 GPUs, in which 
case the training phase was completed 59 minutes. 
The AI models presented in this paper are the top 
classifiers, in terms of ROC and PR analyses, 
from this training campaign. In the context of AI 
models trained to find events in real data, we 
fine-tuned a suite of 20 AI models using H, L and V data. 
This fine-tuning process was completed, for each 
AI model, within 320 minutes using 48 NVIDIA A100 GPUs. We 
then used the same figures of merit described above, 
such as ROC AUC and PR AUC to identify the top six AI 
classifiers that we used for AI inference.

\paragraph{AI inference at scale.} Once we completed the training of multiple AI models 
with optimal classification performance, we 
distributed the inference over a decade of 
simulated gravitational wave data. Leveraging data parallelism during the inference phase allows us to simultaneously distribute data across multiple GPUs, facilitating real-time detection capabilities. This approach not only enhances the throughput of our system by making efficient use of available GPU resources but also significantly reduces latency, ensuring that predictions are generated swiftly and accurately. By parallelizing the data processing workload, we can handle larger volumes of data in real time, maintaining high performance even under heavy load conditions. We first processed 
these data using our AI ensemble using 128 A100 GPUs in 
Polaris, and then post-process the AI models' output 
using 128 CPU nodes in the Theta supercomputer. 
The first part is completed within 3.19 hours, 
while the second is completed in about 0.3 hours. Thus, 
a decade's worth of gravitational wave data 
is processed within 3.5 hours.

\begin{figure*}[!htpt]
    \centering
    \includegraphics[width=\textwidth]{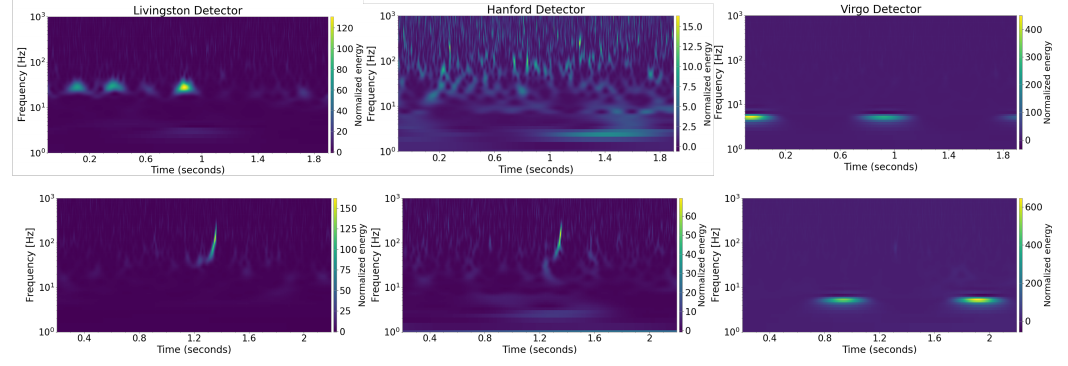}
    \caption{\textbf{Post-processing of noise triggers identified by AI ensemble.} Spectrograms of 
    a noise anomaly at GPS time 1265332743.9816895 (top panel), and the gravitational wave event GW200129\_065458\_1264314069 (bottom panel).}
    \label{fig:event_fps}
\end{figure*}

\noindent \paragraph{AI inference on 
O3b data.} We used an additional step to process real 
data. First we processed the entire month of February 
2020 with our ensemble of 6 AI models. This 
analysis was completed within 1 hour using a single 
NVIDIA V100 GPU. AI inference identified 13 noise 
triggers, which we then followed up using Data Quality 
tools provided by the Gravitational Wave Open Science 
Center, to ensure that the data has the right quality, 
as well as spectrograms to discard clear noise 
anomalies. These lightweight post-processing analyses 
flagged 7 noise anomalies and 6 clean events. A sample 
is shown in Figure~\ref{fig:event_fps}, which presents 
spectrograms of a real event, GW200129\_065458\_1264314069, 
and a noise anomaly located at GPS time 1265332743.9816895.

\section*{Discussion}

We have demonstrated how to encapsulate temporal 
and spatial aspects of 
gravitational wave detection into a novel 
spatiotemporal-graph AI model. We have tested 
this methodology considering a 4D signal manifold 
that describe binary black hole mergers. Our findings 
indicate that AI ensembles can attain optimal 
signal detection while also reducing the 
number of false positives to only 2 over a decade of 
searched data that describes a three detector network. 
This is the first time AI  
attains this level of accuracy and sensitivity 
for gravitational wave detection.

On the algorithm side of things, this work also 
represents a milestone in the training of 
AI models that combine hybrid dilated convolution networks 
and graph neural networks, since we succeeded in 
training these complex models at scale using up to 
256 A100 GPUs with optimal performance, and demonstrated 
strong scaling using up to 512 A100 GPUs.

In addition to reducing time-to-solution using 
distributed training, we also distributed AI inference 
so as to process an entire decade of advanced 
gravitational wave data within 3.5 hours. In brief, 
AI provides a natural pathway to accelerate and 
automate gravitational wave discovery using 
existing supercomputing resources. 

We have explored the use of these 
AI ensembles to search for gravitational waves in O3b 
data available at 
the Gravitational Wave Open Science Center. We did 
this by processing  Advanced LIGO and Advanced Virgo data that span the month of February 2020. 
Our findings indicate that we can indeed identify 
6 HLV events that span the signal manifold used to train 
our AI ensembles, and readily discard noise anomalies. 
This is the first analysis in the literature that 
introduces AI models trained with higher order wave modes 
for an astrophysical population of quasi-circular, 
spinning, non-precessing binary black hole mergers.

Using real data, we identified a 
number of challenges related to the disparate 
sensitivities of the HLV detector network. 
A potential approach for a more resilient AI 
architecture may consist of aggregating 
predictions from the H and L channels (since they have comparable sensitivity) to figure out whether a given signal is concurrently observed by the Advanced LIGO detectors, and then look at the output from Advanced Virgo. This third output may provide additional confidence for a detection 
if any of the Advanced LIGO detectors is down and the signal is also visible in Advanced Virgo. However, AI predictions from advanced Virgo may not be used to discard potential events if they are not clearly extracted from Advanced Virgo data. This condition may be used to design AI models that consider a larger detector network, and may be relaxed once all detectors reach comparable sensitivity.

We also compared the 
performance of our AI ensemble with production 
scale analyses done with established 
gravitational wave pipelines, and then 
conducted a \texttt{PyCBC} search for 10 
gravitational wave sources using available 
gravitational wave data as is. These 
comparisons elucidated the strengths and weaknesses 
of each approach, and indicated that 
AI methods will greatly benefit from using 
information regarding data calibration, 
data quality input, and data cleaning. These 
additional pre-processing analyses should 
be incorporated in future AI analyses.

On the algorithm side of things, 
the choice of pooling operations in the GNN block 
can influence the quality of the learned node representations and 
ultimately affect the performance of our AI models. Currently, we are 
utilizing max pooling based on our preliminary ablation experiments with 
addition, max pooling, and average pooling. The specific nature of signal 
detection seems to favor max pooling since it provides best convergence 
and evaluation performance. However, in future work 
it will be worth exploring the idea of integrating multiple pooling strategies 
and concatenating their outputs. This 
approach may allow the model to capture a more comprehensive set of 
features and potentially generalize better.

This work sets the stage for 
the development of resilient AI approaches for signal detection using open source data at 
the Gravitational Wave Open Science Center to process 
Advanced LIGO and Advanced Virgo data at scale. The next frontier is the 
development of such AI models considering 
higher order wave modes that describe spinning and 
precessing binary black hole mergers.

\section*{Data availability}

We produced datasets of modeled 
waveforms using the \texttt{IMRPhenomXPHM} 
model~\cite{2021PhRvD.103j4056P}, and recolored 
gravitational wave noise using the open 
source \texttt{PyCBC} 
library~\cite{pycbc_library}. The Power 
Spectral Density noise curves for Advanced 
LIGO (aligo\_O4high.txt), 
and Advanced Virgo (avirgo\_O5low\_NEW.txt) are 
readily available at the 
LIGO Document Control Center Portal~\cite{pdsnoise}. 

\section*{Code availability}

The scientific software to reproduce our 
results, including data, AI models for 
synthetic and real noise, postprocessing 
code, and a tutorial for their use is available at 
\texttt{GitHub}~\cite{githubgraph}. 

\section*{References}
\bibliography{references}
\bibliographystyle{iopart-num}


\section*{Acknowledgements}
This work was supported by Laboratory 
Directed Research and Development (LDRD) funding 
from Argonne National Laboratory, provided by the 
Director, Office 
of Science, of the U.S. Department of Energy under 
Contract No. DE-AC02-06CH11357.
E.A.H. and M.T. gratefully acknowledge 
support from National Science Foundation award OAC-1931561 and OAC-2209892. This research used resources of the Argonne 
Leadership Computing 
Facility, which is a DOE Office of Science User 
Facility supported under Contract DE-AC02-06CH11357. 
This research used the Delta advanced computing 
and data resource which is 
supported by the National Science Foundation (award 
OAC 2005572) and the State of Illinois. Delta is a 
joint effort of the University of Illinois at
Urbana-Champaign and its National Center for 
Supercomputing Applications. We thank Marlin Schafer 
for advice on the use of \texttt{IMRPhenomXPHM} for 
waveform production. We thank 
the reviewers for their constructive comments.

\section*{Contributions}
E.A.H. envisioned and led this work, and guided 
the design of AI models. M.T. designed novel 
approaches to develop the neural networks introduced in 
this article, and tested their performance with 
a variety of uncertainty quantification metrics. H.Z. 
distributed the training and inference of AI models 
in the Polaris supercomputer. P.K. 
carried out the \texttt{PyCBC} searches with 
open source data. All authors contributed 
to the writing and reviewing of this manuscript.

\end{document}